# Electrical tuning and switching of an optical frequency comb generated in aluminum nitride microring resonators

**Hojoong Jung, King Y. Fong, Chi Xiong[†], and Hong X. Tang[*]**

*Department of Electrical Engineering, Yale University, New Haven, Connecticut 06511, USA*
[†] *Current address: IBM T.J. Watson Research Center, Yorktown Heights, New York 10598, USA*
*Corresponding author: hong.tang@yale.edu*

Aluminum nitride has been shown to possess both strong Kerr nonlinearity and electro-optic Pockels effect. By combining these two effects, here we demonstrate on-chip reversible on/off switching of the optical frequency comb generated by an aluminum nitride microring resonator. We optimize the design of gating electrodes and the underneath resonator structure to effectively apply electric field without increasing the optical loss. The switching of the comb is monitored by measuring one of the frequency comb peaks while varying the electric field. The controlled comb electro-optic response is investigated for direct comparison with the transient thermal effect.

In a high-quality-(Q)-factor optical cavity, optical frequency comb can be generated by cascaded four-wave mixing (FWM) due to the power enhancement in the cavity with Kerr nonlinearity materials [1,2]. Recently, frequency comb based on silica [3], crystalline $CaF_2$ [4,5] and $MgF_2$ [6,7] microresonators have been reported with tapered optical fiber or prism coupling method that induces minimal insertion loss. Meanwhile, robust on-chip comb generation has been demonstrated in various materials, such as doped silica [8], silicon nitride (SiN) [9-11], and aluminum nitride (AlN) microring resonators [12]. These compact micro-resonators based frequency comb generations have great potential in applications such as precision frequency reference [13], high-speed telecommunications [14], molecular fingerprinting [15], line-by-line pulse shaping [16] and etc.

To make the optical frequency comb more versatile, the capability to dynamically control the comb generation is highly desirable for the purpose of switching and stabilization. It is known that frequency combs can be stabilized thermally by a mechanism known as "thermal lock" [17]. However, the relatively slow response time constant poses a limitation of these control mechanisms. In our previous experiments, we have demonstrated that aluminum nitride microring resonator possesses both strong Kerr nonlinearity which allows generation of optical frequency combs by cascaded FWM [12], and strong Pockels effect which allows GHz electro-optic modulation [18]. By taking advantage of both properties, here we report high speed switching of optical frequency comb in AlN microring resonators. Our system shows promise of building CMOS compatible, high speed, tunable and switchable frequency combs.

Figure 1 (a) shows a top-view optical micrograph of the fabricated device, and Fig. 1 (b) is an SEM image of the cross-section of the AlN waveguide. The AlN is deposited on top of 3 um-thick thermally grown $SiO_2$ on prime Si wafer using S-gun magnetron reactive sputtering system. In order to satisfy the phase matching condition for the frequency comb generation, we engineer the waveguide structure to have a near zero group velocity dispersion (GVD) at 1550nm wavelength. Technical details of the waveguide design can be found in Ref. [12]. The optimized waveguide structure has a top width of 3.5 μm and a height of 650 nm. The 9° angled sidewall is induced by the anisotropic plasma etching process. The waveguide is surrounded by a bottom cladding of a 2 μm thick thermally grown $SiO_2$ and a top cladding of a 3.5 μm thick plasma–enhanced-chemical-vapor-deposited (PECVD) $SiO_2$. To improve the quality of the PECVD $SiO_2$, the device is annealed at 1000 °C for 30 hours in $O_2$ atmosphere. On top of the cladding, an electrode consisting of Ti (10nm) and Au (200nm) is deposited directly over the upper cladding of the waveguide, which is shown in white-yellow color in Fig. 1 (a). The separation between the electrode and the waveguide is designed to be 3.5 μm, which is close enough so that the electric field extending from the electrode can effectively influence the AlN waveguide yet far enough that the optical loss is not increased. Fig. 1 (c) shows the finite element simulation of the static electric potential distribution when 1 V is applied to the electrode (with reference to the bottom of 500 μm thick Si substrate). The effective electric field strength inside the AlN waveguide is around 50 kV/m as shown in Fig. 1 (d). The fractional portion of the optical field of the fundamental TE mode near the edge of the electrode is on the order of $10^{-8}$ which causes a negligible amount of loss in the resonator.

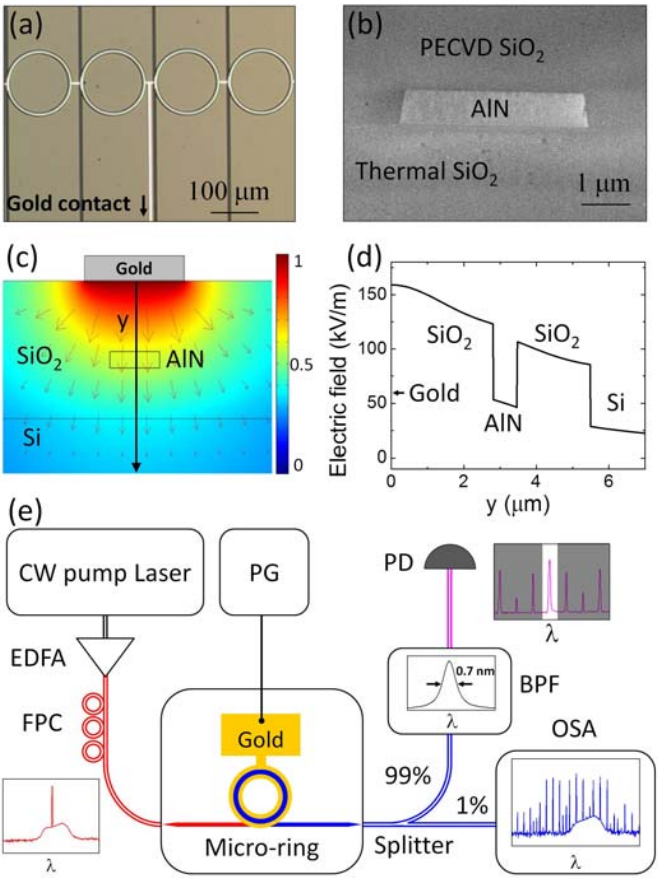

Fig. 1 (a) Optical micrograph showing the top-view of the AlN waveguides, micro-ring resonator and the gold electrodes. (b) SEM image showing the cross-section of the AlN waveguide with width of 3.5 μm and height of 650 nm. (c) FEM simulation of the electric potential distribution when 1V is applied to the gold electrode. (d) Cross-sectional plot of the simulated electric field at the black solid line in (c). (e) The experimental set up for comb generation and modulating. The light before and after the device and the light after going through the bandpass filter are represented by red, blue, and magenta lines, respectively. CW: continuous wave, EDFA: erbium-doped fiber amplifier, FPC: fiber polarization controller, PG: pulse generator, PD: photo detector, BPF: band pass filter, OSA: optical spectrum analyzer.

Figure 1 (e) shows the schematic of the experimental setup. A tunable continuous wave (CW) diode laser amplified up to 3 W by an erbium-doped fiber amplifier (EDFA) is used as the pump (colored as red in Fig. 1 (e)). Before coupled into the chip, a fiber polarization controller (FPC) is used to adjust the laser polarization to be TE-like mode for optimal comb generation performance. A single mode fiber with 4 μm mode-field-diameter (MFD) and an inverted taper coupler on the chip are used for efficient fiber-to-chip coupling, with which we achieve 40% coupling efficiency per facet. Close to the microring resonator, we use a 600 nm wide bus waveguide to couple light into and out from the resonator. The transmitted light is collected by a single mode fiber, 1% of which is sent to optical spectrum analyzer (OSA) for spectral characterization and the remaining is sent through an optical band-pass filter (BPF) with 0.7 nm 3 dB-bandwidth to a photodetector (PD). The tunable BPF is used to pick one of the frequency comb peaks, 1564.5 nm peak in this case, for time domain characterization. The red and blue lines in Fig. 1 (e) represent the light before and after entering the microring resonator and the corresponding spectra are illustrated in the schematics at the lower-left and lower-right of the figure, respectively.

A spectrum of the frequency comb taken by the OSA is shown in Fig. 2 (a). For a microring resonator with radius of 60 μm, the gap between the ring and the waveguide of 500 nm and an optical Q of 500,000 (shown in the upper left inset of Fig. 2 (a)), the optical frequency comb can be generated with 600 mW of input power in the waveguide. The procedure to start the comb generation is described as follows. First, we continuously tune the pump laser wavelength to approach the resonance from the shorter wavelength (blue) side. As the input laser is getting closer to resonance, the circulating power inside the resonator increases, which in turn heats up the resonator causing the resonance wavelength shifting to longer wavelength side (red) due to thermo-optic effect. As the pump wavelength gets closer to resonance, the resonance shifts away accordingly and as a result the transmission profile appears as triangle-shaped, as shown in the upper right inset. During the tuning process when the circulating power inside the resonator reaches a threshold value (shown as the red dotted line in the upper right inset), the stimulated four wave mixing process starts and generates the frequency comb.

To manifest electro-optical tuning of the device, we apply DC voltage to the electrode in the range of -100 to 100 V. As shown in Fig 2 (b), the resonant peaks shift linearly with the applied voltage, having a red (blue) shift of the resonance with positive (negative) voltage. There is no observable change in optical Q due to applied voltage. The plot of resonance wavelength versus applied voltage in Fig. 2 (c) shows that tunability of 0.18 pm/V or 24 MHz/V is achieved. This value is well matched with the result of the finite element simulation (shown Fig. 1 (d)) where the corresponding electric field strength of 45 kV/m (per 1 V applied to the electrode) inside the waveguide is expected.

To demonstrate the frequency comb on/off switching, we apply pulses with an amplitude of 40 V, pulse width of 100 ns and repetition rate of 1 MHz to modulate the resonance condition while monitoring the power of one of the comb peaks selected by the BPF as shown in Fig. 1 (e). Figure 3 (a) shows the applied voltage in time domain and the corresponding comb line response. The inset is a schematic of comb generation with a microring resonator, illustrating the measured comb peak after the BPF. Initially the frequency comb is set to be "on" at zero applied voltage. The monitored comb finger is around 2 mW. When a positive voltage is applied (with fixed pump), the device is switched from "on"-state to "off"-state in around 10 ns. This is limited by the rising time of the high voltage pulse but not the response time of the device, because the photon lifetime in the optical microring cavity is around $\tau = 0.4$ ns ($\tau = Q/2\pi f$, where $Q$ is the quality factor and $f$ is the optical frequency). When the applied

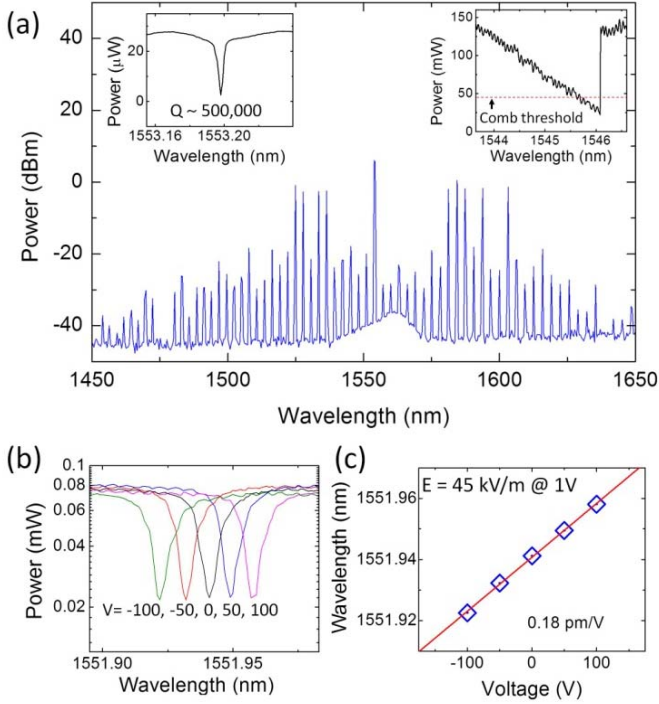

Fig. 2 (a) Optical frequency comb from aluminum nitride ring resonator with 60 µm ring radius. The left inset shows the spectrum of the resonance ($Q \sim 500,000$) used in frequency comb generation. The right inset shows the transmission spectrum of the pump with 600 mW power. The threshold for comb generation is indicated by the dotted line. (b) Spectra of the optical resonance when the DC voltage from -100 to 100 V is applied. (c) Resonant wavelength plotted versus applied voltage and a linear fit. Electric field is estimated to be 45 kV/m when 1V is applied.

voltage is eliminated, the comb is turned on again after an overshoot with 40 ns time constant.

We attribute the overshoot to the thermal-optic effect. Figure 3 (b) illustrates the optical resonance curves at four different time during one cycle from A to D. The vertical purple dashed line represents the fixed pump wavelength, and the four black dots are the corresponding pump output power at four different steps A through D. Note that the closer the pump wavelength is set to the resonant wavelength, the smaller the pump output power and the higher circulating power in the microring. In this case, state D has the highest circulating power while B has the lowest circulating power. If the pump is tuned below red dashed horizontal line in Fig. 3 (b), the circulating power is higher than the threshold power, turning on the comb. Here, black (A) and green (D) curves correspond to comb-on resonant profiles and blue (B) and red (C) curves are comb-off resonant profiles. The blue and red arrows indicate the electric and thermal shifts, respectively. Initially the comb is on at zero applied electric field (A state). Upon applying the electric field, the resonance quickly shifts to the longer wavelength (B) due to electro-optic effect (blue arrow). This in turn decreases the microring circulating power and turns off the comb. However, as the resonator is cooled down as the result of the lowered circulating power, the resonance shifts to the shorter wavelength side and reaches a stable state C (red arrow). In both B and C states, the circulating power at the fixed pump wavelength is lower than threshold power. Therefore the comb is stably off as shown in Fig. 3 (a). When the applied voltage is removed, the resonant peak shifts backward from C to D state electrically (blue arrow). The high circulating power at this point induces not only the overshoot which shows very high nonlinear effect, but also the thermal shift to original state A (red arrow). These two electro-optical and thermal shifts form one on-and-off switching cycle.

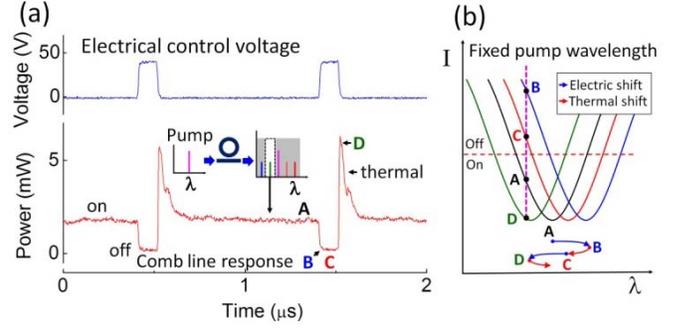

Fig. 3. (a) Applied voltage and the power of the filtered comb peak in time domain. Inset illustrates the spectrum of the comb after BPF. (b) The schematic diagram of optical resonance curves during one cycle from A to D. The red dashed line represents the comb generation threshold and the purple dashed line represents the fixed pump wavelength. Blue and red arrows represent the electrical and thermal shift, respectively.

To further evaluate and separate the electro-optic and thermo-optic effects, we perform measurements in the linear region where the CW laser is set far below threshold power with fixed wavelength near the resonant peak. Figure 4 (a) shows the applied voltage with amplitude of 40 V and pulse width of 500 ns. Since comb is not generated, the through-port transmitted power is directly monitored as voltage is varied (black curve in Fig. 4 (a)). The electro-optical and thermal-optical effects can be clearly identified: the sharp switching corresponds to electro-optic effect whereas the transient decay is due to thermal effect. To understand these feature, in Fig. 4 (b) we show four resonant peaks from A to D, but without threshold line. Similar to the case of Fig. 3 (b), when the applied voltage is turned on, the resonant peak moves from A to B electro-optically (blue arrow). Then the low circulating power pushes back the resonant peak to the steady state C (red arrow). The optical output sharply increases from A to B and then slowly decays from B to C with 40 ns time constant as shown in Fig. 4 (a). Reversely, when the applied voltage is turned off, the device first reaches a transient state D that shows low output power (blue arrow). Then the resonance drifts away to the original stage A due to the high circulating power (red arrow). The overshooting and undershooting output power waveform in Fig. 4 (a) is well matched to the intensity variation at the fixed wavelength in Fig. 4 (b).

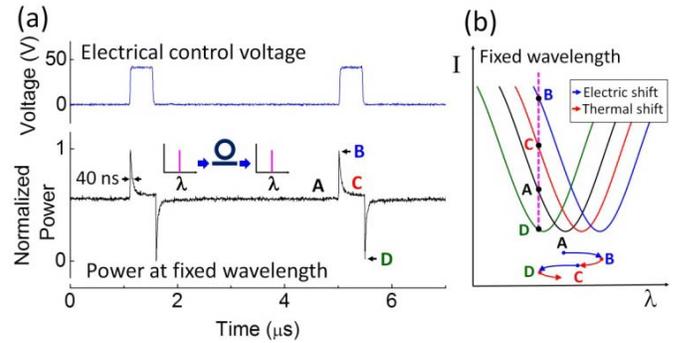

Fig. 4. (a) Applied voltage and the normalized transmission of the output in time domain. Inset illustrates the measured through-port power. (b) The schematic diagram of optical resonance curves during one cycle from A to D. The purple dashed line represents the fixed wavelength.

In conclusion, we demonstrate electro-optical modulation of the optical frequency comb generated from an AlN microring resonator. By applying voltage pulse to the control electrode, high speed frequency comb on/off switching is demonstrated. We also investigate the strong thermo-optic effect due to the high circulating power in the resonator which often plays an important role in nonlinear optics. The influence of the thermal effect can be minimized with a faster modulation than the thermal time constant (~40 ns). With both strong Kerr nonlinearity and electro-optic effect, AlN shows switchable optical frequency comb generation which is promising for applications in spectroscopy, high-speed telecommunications and arbitrary waveform generation.

This work was supported by a Defense Advanced Research Projects Agency (DARPA) grant under its PULSE program. H.X.T. acknowledges support from a Packard Fellowship in Science and Engineering and a National Science Foundation CAREER award. Facilities used were supported by Yale Institute for Nanoscience and Quantum Engineering and NSF MRSEC DMR 1119826. The authors thank Michael Power and Dr. Michael Rooks for assistance in device fabrication.